\documentclass[12pt,a4paper]{article}

\usepackage{amsmath, amssymb, amsfonts}
\usepackage{mathtools}
\usepackage{bm}            
\usepackage{physics}       
\usepackage{siunitx}       

\usepackage{graphicx}
\usepackage{float}
\usepackage{subcaption}
\usepackage{tikz}          
\usepackage{pgfplots}      
\pgfplotsset{compat=newest}

\usepackage[margin=2.5cm]{geometry}
\usepackage{setspace}
\usepackage{hyperref}
\hypersetup{
    colorlinks=true,
    linkcolor=blue,
    citecolor=blue,
    urlcolor=blue
}
\usepackage{cleveref}   

\usepackage[utf8]{inputenc}
\usepackage{amsmath,amssymb}
\usepackage{braket}
\usepackage{physics}
\usepackage{authblk}

\author{C. Abugattas Chacoff}
\affil{\small
    Facultad de Ingeniería, Universidad Alberto Hurtado, Santiago de Chile 8340575, Chile \\ 
    \texttt{cabugattas@uahurtado.cl}
}
\date{} 
\title{A formal correspondence between Bayesian inference problems and the Heisenberg representation}

\begin{document}
\maketitle
\begin{abstract}

The proposed formulation establishes a formal correspondence between the Heisenberg representation and the Bayesian formulation of inverse problems. The likelihood function is first expressed in Hilbert space through a quadratic norm of the difference between the noisy observations and the response of a forward model depending on the unknown parameters \(\boldsymbol{\nu}\), weighted by the inverse of the observational noise covariance. Based on the common mathematical structure of this quadratic form and the observable operator in the Heisenberg representation, the inverse observational noise covariance operator is identified with the initial observable \(\hat{A}(\mathbf{s},0)\equiv\Gamma^{-1}_{\mathrm{noise}}\). To establish this correspondence, a stochastic field with Matérn covariance \(C\) is considered and represented through the Karhunen-Loève expansion. The eigenvalues and eigenfunctions of this expansion are then used to define the state \(|\Psi\rangle\) within the proposed formalism. The main results are summarized by four theorems: the unitarity of the evolution operator, the self-adjointness of the Hamiltonian, the conservation of the trace of the observable, and the formal correspondence between the Bayesian likelihood and the Heisenberg observable structure. Limiting cases of the Matérn covariance are analyzed as \(\ell \to 0\) and \(\ell \to \infty\). Finally, numerical examples illustrate the analytical results.
\end{abstract}

\section{Introduction}
\label{sec:introduccion}

Inverse methods have become an essential tool in different fields of science and engineering, such as geophysics \cite{abugattas_entropy_2026,aster_2018}, biomechanics, materials engineering, and medical tomography \cite{carpio_2023, kaipio_2005}, among others \cite{carpio_2020}. The aim of these methods is to estimate unknown parameters—such as the physical or geometric properties of a medium or anomaly, based on generally noisy observations \cite{kaipio_2005}. There are multiple approaches to addressing inverse problems, from deterministic methods such as the Levenberg-Marquardt minimization method \cite{levenberg_1944, marquardt_1963}, to probabilistic methods such as Bayesian inference \cite{kaipio_2005, tarantola_2005}. In the Bayesian approach, the posterior probability \(f_{post}(\boldsymbol{\nu})\) of the unknown parameter \(\boldsymbol{\nu}\) is obtained using Bayes's theorem by combining the prior and the likelihood probabilities \cite{tarantola_2005}.

In this paper, both the prior probability and the likelihood are assumed to be Gaussian, a choice widely used in the literature \cite{stuart_2010}. In this context, the posterior probability is expressed as an exponential whose argument is the sum of two quadratic terms: one corresponding to the residual between the observed data and the forward model, and the other corresponding to the residual between the estimated parameter and its initial value \cite{tarantola_2005}. The likelihood function is typically expressed in terms of the weighted Euclidean distance between the observed noisy data \(\mathbf{d}^{obs}\) and the data predicted by the model \(\mathbf{f}(\boldsymbol{\nu})\) \cite{stuart_2010, tarantola_2005}.

This formulation is natural in the Hilbert space of square-integrable functions, where the weighted Euclidean norm can be interpreted as the inner product defined by the operator \(\Gamma_{\mathrm{noise}}^{-1}\) \cite{abramowitz_1972}. However, in practice, the observational noise covariance \(\Gamma_{\mathrm{noise}}\) is often unknown, and its estimation constitutes one of the main challenges in Bayesian inference \cite{calvetti_2008}.

This paper proposes a formal correspondence between Bayesian inference and the Heisenberg representation that allows the inverse observational noise covariance operator \(\Gamma^{-1}_{\mathrm{noise}}\) to be interpreted as an observable operator. It is important to note that this work does not quantize a physical system, but rather uses the Heisenberg representation formalism as a mathematical tool to address Bayesian inverse problems. In particular, the identification \(\hat{A}(\mathbf{s},0) \equiv \Gamma_{\mathrm{noise}}^{-1}\) is proposed based on the common mathematical structure of the quadratic likelihood norm and the observable operator in the Heisenberg representation. To establish this correspondence, a stochastic  field \(u(\mathbf{s},t;\alpha)\) is expanded using the Karhunen–Loève expansion, and the state \(|\Psi\rangle\) is constructed from this representation. The observable \(\hat{A}(\mathbf s, \tau)\) evolves over time according to the Heisenberg equation, whose generator is a Hamiltonian constructed from the Matérn kernel.

The main results of this work are summarized in four theorems. \textbf{Theorem 1} establishes the unitary nature of the time-evolution operator. \textbf{Theorem 2} demonstrates the self-adjointness of the Hamiltonian. \textbf{Theorem 3} shows that the trace of the observable \(\hat{A}(\mathbf{s},\tau)\) is invariant under unitary evolution, which implies that the sum of its eigenvalues is conserved over time. \textbf{Theorem 4} establishes the formal correspondence between the quadratic structure of the Bayesian likelihood and the initial observable in the Heisenberg representation. Finally, the limiting cases of the Matérn covariance when the length parameter \(\ell \to 0\) and \(\ell \to \infty\) are analyzed to characterize the proposed formulation under vanishing and long range correlation.

The article is organized as follows. In section \ref{sec:preliminaries}, the necessary mathematical preliminaries are presented, including Hilbert space, Fock space, the Matérn covariance operator, and the Karhunen-Loève expansion. In section \ref{sec:hamiltonian}, the Hamiltonian of the system is constructed in Fock space. In section \ref{sec:psi_state}, the formulation of the state \(|\Psi\rangle\) written as a stochastic field based on the Karhunen–Loève expansion is established. In section \ref{sec:observable_evolu}, the time evolution of an observable in the Heisenberg representation is studied, and its expected value is defined. In section \ref{sec:bayes}, the proposed formalism is linked to Bayesian inference, and the identification \(\hat{A}(\mathbf{s},0) \equiv \Gamma_{\mathrm{noise}}^{-1}\) is established. In section \ref{sec:theorems}, the four theorems are presented along with their respective proofs, and the limit cases of the Matérn covariance are analyzed when \(\ell \to 0\) and \(\ell \to \infty\); in section \ref{sec:numerical_examples} numerical results are presented to illustrate the analytical results. Finally, a discussion of the work is provided in section \ref{sec:conclusions}.

\section{Mathematical prerequisites}
\label{sec:preliminaries}

\subsection{Hilbert Space of functions}
\label{subsec:hilbert}
A Hilbert space is an inner product space—either in the real or complex numbers—that is complete with respect to the norm induced by the inner product \cite{abramowitz_1972, nahakara_geometry}.

In this paper, the Hilbert space of interest is the space of square-integrable functions \(L^2(\Omega)\), where \(\Omega = D \times I\), with \(D \subset \mathbb{R}^N\) the spatial domain and \(I = [0,T]\) the time domain \cite{reed_simon_1980}. The inner product in \(L^2(\Omega)\) is defined as,

\[
\langle f | g \rangle = \int_{\Omega} d\mathbf{s} dt \ f^*(\mathbf{s},t) g(\mathbf{s},t),
\]

and the associated norm is,

\[
\| f \| = \left( \int_{\Omega} d\mathbf{s} dt\ |f(\mathbf{s},t)|^2 \right)^{1/2}.
\]

This space provides the natural framework for the study of random fields and integral operators, such as those appearing in the Karhunen–Loève expansion \cite{ghanem_spanos_2003} and in the definition of quantum observables in this work \cite{sakurai_2020}.

\subsection{Fock space and second quantization}
\label{subsec:fock}

\subsubsection{Fock space}

In quantum mechanics, Fock space is a special Hilbert space constructed as the direct sum of tensor products (symmetrized or antisymmetrized) of another Hilbert space for a single particle \cite{derezinski_qft}. This space is used to describe the quantum state of a system consisting of a variable or indeterminate number of particles.

Fock space is constructed from a basis known as the occupation number; that is, an occupation number basis specifies how many particles are in each possible quantum state \cite{sakurai_2020}. As in Hilbert space, Fock space contains creation operators \(a^{\dagger}_{i}\), which add a particle to state \(i\), and annihilation operators \(a_{i}\), which remove a particle from state \(i\) \cite{fetter_walecka_2003}. These operators satisfy the commutation relations (for bosons) or anticommutation relations (for fermions) \cite{sakurai_2020},

\[
[a_i, a_j^\dagger] = \delta_{ij}, \qquad [a_i, a_j] = 0, \qquad [a_i^\dagger, a_j^\dagger] = 0,
\]

for bosons, and

\[
\{a_i, a_j^\dagger\} = \delta_{ij}, \qquad \{a_i, a_j\} = 0, \qquad \{a_i^\dagger, a_j^\dagger\} = 0,
\]

for fermions.

The Fock space formulation allows the field to be described using creation and annihilation operators associated with the Karhunen–Loève modes.

\subsubsection{Second quantization}

Second quantization is a formulation that allows us to describe systems with a variable number of particles, such as those found in electromagnetic fields, in contrast to first quantization, which deals with systems of individual particles \cite{fetter_walecka_2003, peskin_schroeder_1995}.

In this approach, particles are viewed as excitations of a quantum field \cite{peskin_schroeder_1995}. One of the characteristics of second quantization is that fields are transformed into operators acting on Fock space, which contains all possible states of the system with different numbers of particles; this allows us to describe processes in which particles are created or annihilated. It helps us introduce the treatment of many-particle systems. Second quantization allows us to interpret quantum fields in terms of particles. Each quantum state can be interpreted as a vector in Fock space \cite{attal_fock}. A state can be represented as a superposition of states with an integer number of quanta associated with the field, each having a well-defined energy \cite{derezinski_qft}.

In the context of this work, second quantization allows the modes of the Karhunen-Loève expansion to be represented by creation and annihilation operators. Each eigenfunction \(\phi_i(\mathbf{s},t)\) of the Matérn covariance is associated with an occupation mode, such that the stochastic field \(u(\mathbf{s},t;\alpha)\) can be expressed as a combination of creation and annihilation operators, with spatial dependence \(\mathbf s\), temporal dependence \(t\), and \(\alpha\) as the realizations of the field. This representation is fundamental for constructing the system’s Hamiltonian in Section \ref{sec:hamiltonian}.

\subsection{Matérn kernel}
\label{subsec:matern}

Consider a stochastic field \(u(\mathbf{s},t;\alpha)\), where the spatial dependence is given by \(\mathbf{s} \in \mathbb{R}^N\) and the temporal dependence by \( t > 0\). The Whittle-Matérn stochastic differential equation that generates the field \(u(\mathbf{s},t;\alpha)\) for each realization \(\alpha\) is \cite{lindgren_2011, rue_held_2005},

\begin {equation}
\frac{1}{\sqrt{\beta}}(I-\ell_{s}^{2}\Delta)^{(\eta + \frac{N}{2})/2} (I-\ell_{t}^{2}\frac{\partial^{2}}{\partial t^{2}})^{(\eta_{t} + \frac{1}{2})/2}u(\mathbf{s},t;\alpha) = W(\mathbf{s},t),
\label{eq:edp_white_noise}
\end{equation}

where \(\Delta\) is the Laplacian operator, \( \frac{\partial^{2}}{\partial t^{2}}\) is the second-order time derivative,\(\eta > 0\) is the spatial smoothing factor, \(\eta_{t} > 0\) is the temporal smoothing factor, \(\ell_{s}\) and \(\ell_{t}\) are the spatial and temporal length-scale parameters, respectively. In equation (\ref{eq:edp_white_noise}), \(W(\mathbf{s},t)\) is a Gaussian white noise process with unit variance and mean zero and \(\beta\) is a constant that depends on the parameters of the Matérn kernel.

The spectral density \(S_{u}(\mathbf k, \omega)\) of the field \(u(\mathbf{s},t;\alpha)\) is defined as,
\begin{equation}
S_{u}(\mathbf k, \omega) = \frac{\beta \mathbb E[|\hat{W}(\mathbf k, \omega)|^{2}]}{\left(1 + \ell_{s}^{2}|\mathbf{k}|^2\right)^{\eta + N/2}\left(1 + \ell_{t}^{2}\omega^2\right)^{\eta_{t} + 1/2}},
\label{eq: spectral_field_density}
\end{equation}

where \(\mathbf{k}\) is the wave vector,  \(\omega\) is the time frequency and \([|\hat{W}(\mathbf k, \omega)|^{2}] = 1\). The equation \eqref{eq: spectral_field_density} allows the spectral density to be written as the product of the spatial spectral density \(S_{\mathbf s}=\frac{1}{\left(1 + \ell_{s}^{2}|\mathbf{k}|^2\right)^{\eta + N/2}}\) with the temporal spectral density \(S_{t}=\frac{1}{\left(1 + \ell_{t}^{2}\omega^2\right)^{\eta_{t} + 1/2}}\). Therefore, the spectral density can be written as,

\begin{equation}
S_{u}(\mathbf k, \omega) = {\beta} S_{\mathbf{s}}(\mathbf k) S_{t}(\omega).
\label{eq: spectral_field_density2}
\end{equation}

From this spectral density, the covariance is obtained using the inverse Fourier transform, that is,

\begin{equation}
C(\mathbf{s}, \mathbf{s'}, t, t') = (2\pi)^{-(N+1)}\int_{\mathbb{R}} \int_{\mathbb{R}^N} S_{u}(\mathbf{k}, \omega) \exp(-i\left( \mathbf{k} \cdot \mathbf{s} - \omega (t - t') \right)) d\mathbf{k} d\omega.
\label{eq:inverse_fourier_covariance}
\end{equation}

Taking the inverse Fourier transform in equation (\ref{eq:inverse_fourier_covariance}) which yields a Matérn-type covariance that depends on position and time, expressed as follows,

\begin{equation}
C(\mathbf{s},\mathbf{s'},t,t')= \frac{2^{1-\eta}}{\Gamma(\eta)} \left(\frac{||\mathbf{s}-\mathbf{s'}||}{\ell_{s}}\right)^{\eta} K_{\eta}\left(\frac{||\mathbf{s}-\mathbf{s'}||}{\ell_{s}}\right) \frac{2^{1-\eta_{t}}}{\Gamma(\eta_{t})} \left(\frac{|t-{t'}|}{\ell_{t}}\right)^{\eta_{t}}  K_{\eta_{t}}\left(\frac{|t-{t'}|}{\ell_{t}}\right),
\label{eq:covariance_matern}
\end{equation}

where \(K_{\eta}\) and \(K_{\eta_{t}}\) are modified Bessel functions of the second kind of orders \(\eta\) and \(\eta_{t}\), respectively. 

Using the same smoothing factor for both space and time, that is, \(\eta=\eta_{t}= 0.5\), the covariance \(C(\mathbf{s},\mathbf{s'},t,t')\) in expression (\ref{eq:covariance_matern}) simplifies to a product of exponential terms, that is,

\begin {equation}
C(\mathbf{s},\mathbf{s'},t,t') = \sigma^{2} \exp(-\frac{||\mathbf{s-s'}||}{\ell_{s}})\exp(-\frac{|{t-t'}|}{\ell_{t}}),
\label{eq:cov_matern_exp}
\end{equation}

where \(||\mathbf{s-s'}||\) is the Euclidean distance between points in a coordinate plane and \( \sigma^{2}\) is the variance of the stochastic field. The expression (\ref{eq:cov_matern_exp}) is a covariance that evolves over time due to the function \({e^ {-\frac{|{t-t'}|}{\ell_{t}}}}\), which is evaluated at each time \( t'\). The term \(\sigma^{2}\) represents the marginal variance of the field \(u(\mathbf{s},t;\alpha)\) at a point.

The Whittle-Matérn operator \((I-\ell_{s}^{2}\Delta)^{(\eta + \frac{N}{2})/2} (I-\ell_{t}^{2}\frac{\partial^{2}}{\partial t^{2}})^{(\eta_{t} + \frac{1}{2})/2}\) from \eqref{eq:edp_white_noise}, determines the covariance operator of the stochastic field. Writing this differential operator as \(L\), the covariance operator is given by \(C=\beta L^{-2}\). This property will be used later to construct the Hamiltonian of the system.

\subsection{Karhunen-Loève expansion}
\label{subsec:kl}

Given a random field \(u(\mathbf{s},t;\alpha)\), whose covariance function is given by the expression (\ref{eq:covariance_matern}), it is possible to perform a spectral decomposition, thereby obtaining the eigenfunctions \(\phi_{i}(\mathbf{s},t)\) and the eigenvalues \(\lambda_{i}\). The associated expression is \cite{ghanem_spanos_2003},

\begin {equation}
\int_{\Omega} d\mathbf{s'} d{t'} C(\mathbf{s, s'},t,t') \phi_{i}(\mathbf{s'},t') = \lambda_{i}\phi_{i}(\mathbf{s},t),
\end{equation} 

where \(\Omega=D\times[0,T]\). The covariance operator admits the spectral representation,

\begin{equation}
C= \sum^{\infty}_{i=1} \lambda_i |\phi_{i}\rangle \langle\phi_{i}|.
\end{equation}

The field \(u(\mathbf{s},t;\alpha)\) admits Karhunen–Loève expansion for each realization \(\alpha\) \cite{ghanem_spanos_2003}, that is,

\begin{equation}
u(\mathbf{s},t;\alpha) = \bar{u}(\mathbf{s},t) + \sum^{\infty}_{i=1} \sqrt{\lambda_{i}} \phi_{i}(\mathbf{s},t) \xi_{i}(\alpha),
\label{eq:KL_expantion}
\end{equation}
where \(\bar{u}(\mathbf{s},t)\) is the mean of the stochastic field, and the coefficients \(\xi_{i}(\alpha)\) satisfy \(\mathbb{E}[\xi_i(\alpha)] = 0\)  and \(\mathbb{E}[\xi_i(\alpha)\xi_j(\alpha)] = \delta_{ij}\). 

It is possible to truncate the Karhunen–Loève expansion in (\ref{eq:KL_expantion}) to the first \(M\) terms \cite{ghanem_spanos_2003}. The eigenvalues of the expansion (\ref{eq:KL_expantion}) are ordered in descending order, such that \(\lambda_1 \ge \lambda_2 \ge \dots \ge \lambda_i \ge \dots\); the leading modes capture the dominant contribution to the variance of the field. The truncation error decreases as $M$ increases, allowing the expansion to be approximated to the first \(M\) terms.

The Karhunen-Loève eigenfunctions \(\phi_{i}(\mathbf s, t)\) diagonalize the covariance operator \(C\) and provide the spectral basis used to represent the stochastic field. 

Once a representation of the field is obtained through the Karhunen-Loève expansion, each mode can be interpreted as an independent degree of freedom. This modal representation allows us to introduce the second-quantization formalism, where the modes are described by creation and annihilation operators in Fock space. The Hamiltonian of the system is then constructed as the energy operator associated with these modes in the following section.

\section{The Hamiltonian}
\label{sec:hamiltonian}
For a system with a variable number of particles, the Hamiltonian is constructed on Fock space through the direct sum of symmetrized or antisymmetrized tensor products of another Hilbert space \cite{fetter_walecka_2003, salcedo_qft}, that is,

\begin {equation}
\hat{H} = \sum^{N}_{i}\hat{H}_i.
\label{eq:sum_hamiltonians}
\end{equation}

In this space, the total Hamiltonian is the sum of the Hamiltonians of each subsystem \(i\), each of which acts trivially on the others. Thus, the total Hamiltonian of the system in its continuous form, in Fock space, and in its second quantization is given by \cite{momeni_fock, salcedo_qft},

\begin{equation}
\hat{H} = \int dt d^{N}\mathbf{s} \hat{\psi}^{\dagger}(\mathbf{s},t) h(\mathbf{s},t) \hat{\psi}(\mathbf{s},t),
\label{eq:total_hamiltonian}
\end{equation}
where \( h(\mathbf{s},t)\) is the single particle operator acting on the creation and annihilation field operators $\hat{\psi}^\dagger(\mathbf{s},t)$ and $\hat{\psi}(\mathbf{s},t)$, respectively. This expression shows that the total Hamiltonian is the sum of the local contributions of the energy density at each point in space.

In expression \eqref{eq:total_hamiltonian}, \( h(\mathbf{s},t) \) is defined as the single-particle operator, that is,

\begin {equation}
h(\mathbf{s},t) = \beta (I-\ell_{s}^{2}\Delta)^{-(\eta + \frac{N}{2})} (I-\ell_{t}^{2}\frac{\partial^{2}}{\partial t^{2}})^{-(\eta_{t} + \frac{1}{2})} = \beta L^{-2},
\label{eq:single-particle_operator}
\end{equation}

where the scaling factor \(\beta\) is included in the definition of
\(h(\mathbf{s},t)\). According to the representation of stochastic partial differential equations \cite{lindgren_2011}, this operator corresponds to the Matérn covariance operator. Therefore, in the present formulation \( h(\mathbf{s},t) \equiv C\).

The eigenfunctions \(\phi_i(\mathbf{s},t)\) are eigenfunctions of both the covariance operator \(C\) and the operator \(h(\mathbf s, t)\). This relation is fundamental for connecting the Karhunen–Loève decomposition with the system’s Hamiltonian.

In this framework, the second-quantization field operators are defined as \cite{momeni_fock},

\begin{align}
\hat{\psi}(\mathbf{s},t) &= \sum_i \phi_i(\mathbf{s},t) \ a_i,
\label{eq:annihilation_operator}\\
\hat{\psi}^\dagger(\mathbf{s},t) &= \sum_i \phi_i^*(\mathbf{s},t) \ a_i^\dagger,
\label{eq:field_operators}
\end{align}

where \(\hat{\psi}(\mathbf{s},t)\) and \(\hat{\psi}^{\dagger}(\mathbf{s},t)\) are the annihilation and creation operators, respectively, and \(\phi_{i}(\mathbf{s},t)\), \(\phi_{i}^{*}(\mathbf{s},t)\) are the eigenfunctions and their conjugate, respectively. In expressions \eqref{eq:annihilation_operator} and (\ref{eq:field_operators}), \(a_i\) and \(a^{\dagger}_{i}\) are the annihilation and creation operators, in a discrete basis.

\section{\texorpdfstring{Definition of the state $|\Psi\rangle$}{Definition of the psi state}}
\label{sec:psi_state}

Based on the Karhunen-Loève expansion, the state \(|\Psi\rangle\) within the proposed formalism depends on the superposition of the eigenvalues of the Matérn covariance with the random coefficients \(\xi_{i}(\alpha)\), so \(|\Psi\rangle\) depends on each realization of the field, that is,

\begin {equation}
|\Psi\rangle = \sum_{i=1}^{\infty} \sqrt{{\lambda_{i}}} \xi_i(\alpha)\ |\phi_{i}(\mathbf{s},t)\rangle,
\label{eq:def_psi_state}
\end{equation}

where \(\{|\phi_{i}\rangle\}\) is an orthonormal basis of the Hilbert space \(\mathcal{H}\), and the coefficients \(\xi_i(\alpha)\) are random variables with \(\mathbb{E}[\xi_i(\alpha)] = 0\) and \(\mathbb{E}[\xi_i(\alpha)\xi_j(\alpha)] = \delta_{ij}\). This identification for the state \(|\Psi\rangle\) is natural, since a stochastic field \(u(\mathbf{s},t;\alpha)\) is given by the KL expansion \eqref{eq:KL_expantion}.

The corresponding space-time representation of the state is obtained by projection onto position-time basis, that is, 
\begin{equation}
\Psi(\mathbf s,t; \alpha) = \langle \mathbf s, t| \Psi(\alpha)\rangle = \sum_{i=1}^{\infty} \sqrt{\lambda_i} \ \xi_i (\alpha)\ \phi_i(\mathbf{s},t).
\label{eq:funcion_onda_KL}
\end{equation}

Both expressions represent the state associated with the stochastic field: the first is the representation of the state in Hilbert space, and the second is its expansion in the basis of Matérn covariance eigenfunctions. 

Now that the state definition \(|\Psi\rangle\) has been written, it is possible to rewrite the total Hamiltonian. To do this, the definitions of the operators \(\hat \psi(\mathbf s,t)\) and \(\hat \psi^{\dagger}(\mathbf s,t)\) are substituted into (\ref{eq:total_hamiltonian}), obtaining the following relation,

\begin {equation}
\hat{H} = \sum_{i,j} \left( \int d\mathbf{s} \ dt \ \phi_i^*(\mathbf{s},t) \ h(\mathbf{s},t) \ \phi_j(\mathbf{s},t) \right) a_i^\dagger a_j.
\label{eq:hamiltoniano_base1}
\end{equation}

Since the functions \(\phi_i\) are the eigenfunctions of the Matérn kernel, the matrix element in parentheses reduces to,

\[
\int d\mathbf{s} \ dt \ \phi_i^*(\mathbf{s},t) \ h(\mathbf{s},t) \ \phi_j(\mathbf{s},t) = {\lambda_j}\delta_{ij},
\]
where \(h(\mathbf s, t)\) is identified with the covariance operator.

Therefore, the Hamiltonian in the eigenfunction basis is given by:

\begin{equation}
\hat{H} = \sum_{i,j} {\lambda_j}\delta_{ij}a_i^\dagger a_j.
\label{eq: hamiltoniano_base_phi}
\end{equation}

\section{The observable and its evolution}
\label{sec:observable_evolu}

In quantum mechanics, the time evolution of an observable $\hat{A}(\mathbf s, \tau)$ in the Heisenberg representation is governed by the equation of motion \cite{sakurai_2020},

\begin {equation}
\frac{d\hat{A}(\mathbf s, \tau)}{d\tau} = -\frac{i}{\hbar}[\hat{A}(\mathbf s, \tau),\hat{H}] + \frac{\partial \hat{A}(\mathbf s, \tau)}{\partial \tau}
\label{eq:heisenberg_solution}
\end{equation}
where \(\hat{H}\) is the Hamiltonian of a system, as seen in the previous section. The formal solution to expression (\ref{eq:heisenberg_solution}) is \cite{peskin_schroeder_1995},

\begin{equation}
\hat{A}(\mathbf s, \tau) = U^\dagger(\tau) \ \hat{A}(\mathbf s, 0) \ U(\tau),
\label{eq:heisenberg_solution_formal}
\end{equation}

where \(U(\tau)\) is the time-evolution operator, expressed as time-ordered exponential as \cite{sakurai_2020},

\begin{equation}
U(\tau) = \mathcal{T} \exp\left( -\frac{i}{\hbar} \int_0^\tau \hat{H}(\tau') d\tau' \right),
\label{eq:ev_temporal_dyson}
\end{equation}

and \(\mathcal{T}\) is the time-ordering operator.

\textbf{Special case:} If the Hamiltonian is time-independent, the evolution operator reduces to \(U(\tau) = \exp(-i\hat{H}\tau/\hbar)\) (or \(\exp(i\hat{H}\tau/\hbar)\) depending on the sign convention), and equation (\ref{eq:heisenberg_solution_formal}) simplifies to the expression,

\[
\hat{A}(\mathbf s, \tau) = \exp(i\hat{H}\tau/\hbar) \ \hat{A}(\mathbf s, 0) \ \exp(-i\hat{H}\tau/\hbar).
\]

In this work, the Hamiltonian has no explicit dependence on the Heisenberg evolution parameter \(\tau\) and the time-ordered evolution operator reduces to \(U(\tau) = \exp(\frac{-i\hat{H}\tau}{\hbar})\).

\subsection{Expected value of a quantum observable}
\label{subsec:expected_value}
The expected value of an observable \(\hat{A}(\mathbf{s},\tau)\) in a state \(|\Psi\rangle\) \cite{nahakara_geometry, sakurai_2020}, is defined as,

\begin {equation}
\langle \hat{A}(\mathbf{s},\tau)\rangle_{\Psi} = \langle \Psi | \hat{A}(\mathbf{s},\tau)| \Psi \rangle,
\label{eq:expected_value}
\end{equation}
where \(\langle \Psi|\) is the hermitian conjugate of the state \(|\Psi \rangle\). The definition in equation (\ref{eq:expected_value}) holds regardless of whether the observable is time-dependent or not.

Using the definition of the observable operator \eqref{eq:heisenberg_solution_formal}, the expected value of \(\hat{A}(\mathbf{s},\tau)\) can be written as follows,

\begin {equation}
\langle \hat{A}(\mathbf{s},\tau)\rangle_{\Psi} = \langle \Psi| U^{\dagger}(\tau) \hat{A}(\mathbf{s},0) U(\tau)| \Psi \rangle.
\label{eq:heisenberg_expected_value}
\end{equation}

Equivalently, introducing the evolved state \(|\Psi(\tau)\rangle=U(\tau)|\Psi\rangle\), the expected value can be expressed as the quadratic form weighted by the initial observable \(\hat{A}(\mathbf{s},0)\) \cite{nahakara_geometry},

\begin {equation}
\langle \hat{A}(\mathbf{s},\tau) \rangle_{\Psi} = \| \Psi(\tau) \|_{\hat{A}(\mathbf{s},0)}^2 = \langle \Psi(\tau) | \hat{A}(\mathbf{s},0) | \Psi(\tau) \rangle,
\label{eq:quantum_quadratic_norm}
\end{equation}

where the quadratic norm weighted by a generic operator \(P\) is defined as \cite{janson_distance, nahakara_geometry},

\[
\| \Psi \|_{P}^2 := \langle \Psi | P | \Psi \rangle,
\]

then \(P=\hat{A}(\mathbf{s},0)\) is identified. This representation is particularly useful when \(\hat{A}(\mathbf{s},0)\) is an inverse covariance operator associated with observational noise.

In the context of stochastic fields, assuming that \(\Psi\) is expanded in a Karhunen–Loève basis, then it is possible to write this state using expression \eqref{eq:KL_expantion} for each realization \(\alpha\), that is, \( \Psi= \sum_{i} \xi_{i}(\alpha)\sqrt{{\lambda_{i}}}|\phi_{i}\rangle\).

Therefore, writing the expected value of \(\langle \hat{A}(\mathbf{s},\tau) \rangle_{\Psi}\) in terms of the Karhunen–Loève expansion (\ref{eq:KL_expantion}) and \eqref{eq:expected_value},

\[
\langle \hat{A}(\mathbf{s},\tau) \rangle_{\Psi} = \sum_{i,j} \sqrt{\lambda_i^{*}}\sqrt{\lambda_j} \ \xi_{i}(\alpha)^{*} \xi_{j}(\alpha) \langle \phi_{i}(\mathbf{s},t) | U(\tau)^{\dagger} \hat{A}(\mathbf{s},0) U(\tau)| \phi_{j}(\mathbf{s},t) \rangle,
\]

where \(\phi_i(\mathbf{s},t)\) and \(\phi_j(\mathbf{s},t)\) are orthonormal bases of the Hilbert space; therefore, \(\langle \phi_i(\mathbf{s},t) | \phi_j(\mathbf{s},t) \rangle = \delta_{ij}\).

Next, the expected value formalism is applied to a stochastic field with Matérn-type spatio-temporal covariance within a Bayesian framework.

\section{Bayesian representation in the Heisenberg formalism}
\label{sec:bayes}

From measurement, the noisy observational data \(\mathbf{d}^{obs}\) are obtained, with an associated error \(\boldsymbol{\varepsilon} \sim \mathcal{N}(0, \Gamma_{\mathrm{noise}})\), where \(\Gamma_{\mathrm{noise}}\) is the covariance matrix of the observational noise and \(\mathbf{f}(\boldsymbol{\nu})\) is the forward model that maps the parameters \(\boldsymbol{\nu}\) to the spatio-temporal data. 

Considering this, the likelihood of the data given the parameters is expressed as \cite{abugattas_quantifying_2025, carpio_2020, carpio_2023},

\begin{equation}
\mathcal{L}(\boldsymbol{\nu} \mid \mathbf{d}^{obs}) \propto \exp\left\{ -\frac{1}{2} \left\| \mathbf{d}^{obs} - \mathbf{f}(\boldsymbol{\nu}) \right\|_{\Gamma_{\mathrm{noise}}^{-1}}^2 \right\},
\label{eq:likelihood}
\end{equation}
where the quadratic norm in the exponent is weighted by the inverse of the noise covariance, that is,\cite{carpio_2023}

\begin{equation}
\mathcal{L}(\boldsymbol{\nu} \mid \mathbf{d}^{obs}) \propto \exp\left\{ -\frac{1} {2}   (\mathbf{d}^{obs} - \mathbf{f}(\boldsymbol{\nu}))^{T}\Gamma_{\mathrm{noise}}^{-1}(\mathbf{d}^{obs} - \mathbf{f}(\boldsymbol{\nu}))  \right\},
\label{eq:likelihood_decomposition}
\end{equation}

To establish the correspondence with the Heisenberg representation, the observational data and the forward-model field are represented in the same Karhunen-Loève basis (introduced in section~\ref{subsec:kl}). In this representation, let \(d_i^{obs}\) and \(f_i(\boldsymbol{\nu})\) denote the KL coefficients obtained by projecting \(\mathbf{d}^{obs}\) and \(\mathbf{f}(\boldsymbol{\nu})\), respectively, onto the eigenfunctions \(\phi_i\), 

\begin{equation}
|\Psi^{obs}\rangle = \sum_i d_i^{obs}|\phi_i\rangle, \qquad |\Psi(\boldsymbol{\nu})\rangle = \sum_i f_i(\boldsymbol{\nu})|\phi_i\rangle.
\label{eq:obs_model_states}
\end{equation}

The residual state is then defined as, 
\begin{equation}
|\Psi_r(\boldsymbol{\nu})\rangle = |\Psi^{obs}\rangle-|\Psi(\boldsymbol{\nu})\rangle = \sum_i \left[d_i^{obs}-f_i(\boldsymbol{\nu})\right] |\phi_i\rangle.
\label{eq:residual_state}
\end{equation}

The KL eigenfunctions are determined by the Matérn covariance operator \(C\) of the stochastic field. However, the residual between the observational data and the forward model, defined as \(r = d^{obs} - f(\mathbf {\nu})\), is weighted by the observational noise covariance \(\Gamma_{\mathrm{noise}}\) in the likelihood probability. These two covariance operators play different roles in this formulation. 

In this work, as a modeling hypothesis, the initial observable \(\hat A(\mathbf s, 0)\) is identified with the inverse observational noise covariance operator, that is,

\begin{equation}
\hat{A}(\mathbf{s},0)\equiv\Gamma_{\mathrm{noise}}^{-1},
\label{eq:A0_noise}
\end{equation}

this identification associates the initial observable with the inverse observational noise covariance operator, and establishes the proposed bridge between the Bayesian formulation and the Heisenberg representation.

Consequently, the quadratic term appearing in the likelihood can be expressed in the proposed Hilbert space, that is (see in appendix \ref{app:appendixA}),

\begin{equation}
(\mathbf{d}^{obs}-\mathbf{f}(\boldsymbol{\nu}))^{T} \Gamma_{\mathrm{noise}}^{-1}(\mathbf{d}^{obs}-\mathbf{f}(\boldsymbol{\nu})) = \langle\Psi_{r}(\boldsymbol{\nu})|\hat{A}(\mathbf{s},0) |\Psi_r(\boldsymbol{\nu})\rangle,
\label{eq:likelihood_observable_equivalence}
\end{equation}

accordingly, the likelihood can be written as,

\begin{equation}
\mathcal{L}(\boldsymbol{\nu} \mid \mathbf{d}^{obs}) \propto \exp\left\{ -\frac{1}{2} \langle\Psi_r(\boldsymbol{\nu})| \hat{A}(\mathbf{s},0) |\Psi_r(\boldsymbol{\nu})\rangle \right\}.
\label{eq:likelihood_heisenberg}
\end{equation}

Bayes' theorem establishes that the posterior probability can be obtained by multiplying the prior probability by the likelihood, that is \cite{Bishop2006, Jaynes2003},

\begin{equation}
f_{post}(\boldsymbol{\nu}) \propto \mathcal{L}(\boldsymbol{\nu} \mid \mathbf{d}^{obs}) \ \mathcal{L}_{prior}(\boldsymbol{\nu}),
\label{eq:bayes_theo}
\end{equation}

where the prior probability is considered as Gaussian, with mean \(\boldsymbol{\nu}_0\) and covariance \(\Gamma_{prior}\) \cite{carpio_2020},

\begin{equation}
\mathcal{L}_{prior}(\boldsymbol{\nu}) \propto \exp\left\{-\frac{1}{2} \left\|\boldsymbol{\nu}-\boldsymbol{\nu_0} \right\|_{\Gamma_{prior}^{-1}}^2\right\}.
\label{eq:prob_priori}
\end{equation}

Therefore, the posterior probability is given by \cite{tarantola_2005},

\begin{equation}
f_{post}(\boldsymbol{\nu}) \propto \exp\left\{-\frac{1}{2}(\mathbf{d}^{obs}-\mathbf{f}(\boldsymbol{\nu}))^{T} \Gamma_{\mathrm{noise}}^{-1}(\mathbf{d}^{obs}-\mathbf{f}(\boldsymbol{\nu})) -\frac{1}{2}(\boldsymbol{\nu}-\boldsymbol{\nu_0})^{T} \Gamma_{prior}^{-1}(\boldsymbol{\nu}-\boldsymbol{\nu_0}) \right\}.
\label{eq:posterior_prob}
\end{equation}

Comparing the residual state defined in \eqref{eq:residual_state}, and the identification \(\hat{A}(\mathbf{s},0)\equiv\Gamma_{\mathrm{noise}}^{-1}\), the quadratic term of the likelihood can be equivalently represented in the KL basis as \(\langle\Psi_r(\boldsymbol{\nu})|\hat{A}(\mathbf{s},0)|\Psi_r(\boldsymbol{\nu})\rangle\). The posterior probability can be equivalently written as, 

\begin{equation}
f_{post}(\boldsymbol{\nu}) \propto \exp\left\{ -\frac{1}{2}\langle\Psi_r(\boldsymbol{\nu})|\hat{A}(\mathbf{s},0) |\Psi_r(\boldsymbol{\nu})\rangle -\frac{1}{2}(\boldsymbol{\nu}-\boldsymbol{\nu_0})^{T} \Gamma_{prior}^{-1}(\boldsymbol{\nu}-\boldsymbol{\nu_0}) \right\}.
\label{eq:prob_posterior_heisenberg}
\end{equation}

In the expression \eqref{eq:prob_posterior_heisenberg} the correspondence is established through the quadratic structure of the Bayesian likelihood, that is, the residual between the observational data and the forward-model response, is represented as a state in the KL basis, while the inverse observational noise covariance operator is identified with the initial observable in the Heisenberg representation. For more details about the equivalence of the Bayesian quadratic form under its representation in the KL basis, see in appendix \ref{app:appendixA}.

\section{Theorems and limit cases}
\label{sec:theorems}

The following theorems establish the fundamental properties of the framework: the unitarity of time evolution, the self-adjointness of the Hamiltonian, the conservation of the trace of the observable, and the formal correspondence between the quadratic norm of the Bayesian likelihood and the initial observable in the Heisenberg representation. After the theorems, the limiting cases of the Matérn covariance are examined, which provide physical insight into the behavior of the field as the correlation length becomes very small or very large.

\subsection{\textbf{Theorem 1: Unitary evolution}}
\label{subsec:theorem_1}

\textit{The time-evolution operator expressed as a time-ordered exponential,}

\textit{\[
U(\tau) = \mathcal{T} \exp\left(
-\frac{i}{\hbar} \int_0^\tau \hat H \ d\tau'
\right)
\]}

\textit{is unitary, that is,}

\textit{\[
U(\tau) U^\dagger(\tau)
=
U^\dagger(\tau) U(\tau)
=
\hat{I},
\]}

\textit{where \(\mathcal{T}\) is the time-ordering operator.}

\subsubsection{Proof}

The Hamiltonian defined in equation \eqref{eq:total_hamiltonian} is given by,
\[
\hat{H}
=
\int dt\ d^{N}\mathbf{s}\
\hat{\psi}^\dagger(\mathbf{s},t)
h(\mathbf{s},t)
\hat{\psi}(\mathbf{s},t),
\]

where $t$ denotes the temporal coordinate of the spatio-temporal stochastic field, whereas $\tau$ denotes the evolution parameter in the Heisenberg representation. Since the total Hamiltonian has no explicit dependence on the $\tau$ variable, it follows that,

\[
\frac{\partial \hat H}{\partial \tau}=0.
\]

The time-evolution operator admits a Dyson expansion, that is,
\[
U(\tau) = \sum_{n=0}^{\infty} \left(-\frac{i}{\hbar}\right)^n \int_0^\tau d\tau_1 \int_0^{\tau_1} d\tau_2 \cdots \int_0^{\tau_{n-1}} d\tau_n\ \hat H^n, \]

where the term for $n=0$ is $\hat I$. Since $\hat H$ does not depend explicitly on $\tau$, it can be taken outside the evolution time integrals. Furthermore, the integrals of the above expression satisfy the next relation,

\[ \int_0^\tau d\tau_1 \int_0^{\tau_1} d\tau_2 \cdots \int_0^{\tau_{n-1}} d\tau_n = \frac{\tau^n}{n!}.\]

Therefore, the Dyson expansion reduces to

\[U(\tau) = \sum_{n=0}^{\infty} \frac{1}{n!} \left(-\frac{i\tau}{\hbar}\right)^n \hat H^n.\]

Since $\hat H$ is self-adjoint, $\hat H^\dagger=\hat H$ (see Theorem 2), the adjoint is,

\[U^\dagger(\tau) = \sum_{m=0}^{\infty} \frac{1}{m!} \left(\frac{i\tau}{\hbar}\right)^m \hat H^m.\]

Multiplying the series expansion of \(U(\tau)\) and \(U^{\dagger}(\tau)\), it follows,

\begin{equation}
\begin{split}
U U^\dagger =& \sum_{n,m=0}^{\infty} \frac{1}{n!m!} \left(-\frac{i\tau}{\hbar}\right)^n \left(\frac{i\tau}{\hbar}\right)^m
\hat H^{n+m}.
\label{eq:u_udagger}
\end{split}
\end{equation}

Rearranging the expression \eqref{eq:u_udagger}  by combining the prefactors and defining $N=n+m$, it follows,

\begin{equation}
U U^\dagger
=
\sum_{N=0}^{\infty}
\left(\frac{i\tau}{\hbar}\right)^N
\hat H^N
\sum_{n=0}^{N}
\frac{(-1)^n}{n!(N-n)!}.
\end{equation}

Using the factorial definition, that is,

\[
\frac{1}{n!(N-n)!}
=
\frac{1}{N!}\binom{N}{n},
\]

the product of the time-evolution operator and its conjugate is written as the following,

\begin{equation}
U U^\dagger = \sum_{N=0}^{\infty} \frac{1}{N!} \left(\frac{i\tau}{\hbar}\right)^N \hat H^N \sum_{n=0}^{N} \binom{N}{n}(-1)^n.
\label{eq:time-evolution-operator}
\end{equation}

The sum over $n$ is evaluated using Newton's binomial theorem,

\[
\begin{aligned}
\sum_{n=0}^{N} \binom{N}{n}(-1)^n &=\sum_{n=0}^{N} \binom{N}{n}1^{N-n}(-1)^n &=(1-1)^N,
\end{aligned}
\]

for any $N>0$, this expression vanishes, whereas for $N=0$ it is equal
to one. Consequently, the only term that survives is the one of order
$N=0$. Therefore,

\[
U U^\dagger=\hat I.
\]

Applying the same procedure to $U^\dagger U$ yields the same result. Thus, $U$ is a unitary operator $\square$.

\subsection{Theorem 2: Self-adjointness of the Hamiltonian}
\label{subsec:teo_2}

\textit{The total Hamiltonian of the system, \(\hat{H}\), is self-adjoint, that is, the following holds:}
\textit{\[ \hat{H}= \hat{H}^{\dagger},\]}
\textit{where \(\hat{H}^{\dagger}\) is the conjugate transpose of \(\hat{H}\).}

\subsubsection{Proof}

The self-adjointness of the Hamiltonian is a crucial property in quantum mechanics; ensures that its eigenvalues are real and that the time evolution is physically consistent, so the goal in this Theorem is to prove that \( \hat{H}= \hat{H}^{\dagger}.\)

Using the definitions of the field operators
$\hat{\psi}(\mathbf{s},t)$ and $\hat{\psi}^{\dagger}(\mathbf{s},t)$, the total Hamiltonian can be rewritten by substituting \eqref{eq:annihilation_operator} and \eqref{eq:field_operators}, which yields the following relation,

\begin{equation}
\hat{H} = \sum_{i,j} \left( \int d\mathbf{s} \ dt \ \phi_i^*(\mathbf{s},t) \ h(\mathbf{s},t) \ \phi_j(\mathbf{s},t) \right) a_i^\dagger a_j.
\label{eq:base_hamiltonian}
\end{equation}

Since the functions \(\phi_i\) are eigenfunctions of the Matérn covariance operator $C$, they are also eigenfunctions of the operator \(h(\mathbf{s}, t)\), so \(h(\mathbf{s},t) \phi_j(\mathbf{s},t) = \lambda_j \phi_j(\mathbf{s},t)\), as defined previously in \ref{sec:hamiltonian}. By the orthonormality of the eigenfunctions, it can be written,

\[
\int d\mathbf{s} \ dt \ \phi_i^*(\mathbf{s},t) \ h(\mathbf{s},t) \ \phi_j(\mathbf{s},t) = \lambda_j \delta_{i,j}.
\]

Therefore, the Hamiltonian in the eigenfunction basis is given by,
\begin{equation}
\hat{H} = \sum_{i,j} \lambda_j \delta_{i,j} \ a_i^\dagger a_j,
\label{eq: hamiltoniano_base_phi2}
\end{equation}

where the Hamiltonian in equation (\ref{eq: hamiltoniano_base_phi2}) involves a sum over two indices. Taking the conjugate of (\ref{eq: hamiltoniano_base_phi2}) yields,

\begin{equation}
\hat{H}^{\dagger} = \sum_{i,j}  \lambda^{*}_{j} \delta_{i,j}\left(a_i^\dagger a_j\right)^\dagger,
\label{eq:transposed_conjugate_Hamiltonian}
\end{equation}

where \( \lambda_j = \lambda^{*}_{j}\) because the eigenvalues are real.

From expression (\ref{eq:transposed_conjugate_Hamiltonian}) and using the adjoint property \cite{sakurai_2020}, 
\[
(AB)^\dagger = B^\dagger A^\dagger,
\] 

together with the properties of creation and annihilation operators, the identities \((a^{\dagger}_i)^{\dagger} = a_{i}\) and \((a_{j})^{\dagger} = a^{\dagger}_{j}\), it can be seen that expression (\ref{eq:transposed_conjugate_Hamiltonian}) takes the following form,

\begin{equation}
\hat{H}^{\dagger} = \sum_{i,j} \lambda_{j}\delta_{i,j}\ a_j^{\dagger} a_i.
\label{eq:hamiltonian_dagger}
\end{equation}

Since the Kronecker delta imposes \(i=j\), the expression \eqref{eq:hamiltonian_dagger} reduces to, 

\begin{equation}
\hat{H}^{\dagger} = \sum_{i} \lambda_{i}\ a_i^{\dagger} a_i.
\end{equation}

On the other hand, the Hamiltonian \(\hat{H}\) when \(i=j\) the Kronecker delta \(\delta_{i,i} =1\), gives,

\begin{equation}
\hat{H} = \sum_{i} \lambda_i \ a_i^\dagger a_i.
\end{equation}

Therefore, \(\hat{H} = \hat{H}^{\dagger}\), proving that the Hamiltonian \(\hat{H}\) is self-adjoint $\square$.

\subsection{Theorem 3: Conservation of the trace of the observable}
\label{subsec:teo_3}

\textit{The trace of the observable \(\hat{A}(\mathbf{s},\tau)\) is invariant under unitary evolution:}
\[
\operatorname{Tr}[\hat{A}(\mathbf{s},\tau)] = \operatorname{Tr}[\hat{A}(\mathbf{s},0)].
\]

\subsubsection{Proof}

By setting \( \hat{A}(\mathbf{s},\tau) = {U}^{\dagger} \hat{A}(\mathbf{s},0) {U} \) \cite{horn_johnson_2013}, the following trace identity is obtained,

\[ \operatorname{Tr}[{U}^{\dagger} \hat{A}(\mathbf{s},0) {U}] = \operatorname{Tr}[\hat{A}(\mathbf{s},0)].\]

Using the cyclic property of the trace \(\operatorname{Tr}[\hat{B}\hat{D}\hat{E}] = \operatorname{Tr}[\hat{E}\hat{B}\hat{D}]\) in the above expression,

\[ \operatorname{Tr}[\hat{A}(\mathbf{s},0){U}{U}^{\dagger}] = \operatorname{Tr}[\hat{A}(\mathbf{s},0)], \] 

since the evolution operator \(U\) is unitary, \(U{U}^{\dagger} = \hat{I}\) (see Theorem 1). Therefore, the following equality holds,

\begin{equation}
\operatorname{Tr}[\hat{A}(\mathbf{s},\tau)] = \operatorname{Tr}[\hat{A}(\mathbf{s},0)],
\label{eq:trace_invariance}
\end{equation}

which demonstrates, via equation (\ref{eq:trace_invariance}), that the trace of the observable \(\hat{A}(\mathbf{s},\tau)\) is invariant under unitary transformations.

Independently, the trace of the covariance Matérn $C$ can be related to the spectral representation through the Karhunen-Loève expansion. Considering a truncation to \(M\) modes, the expected value of the square of the field, expanded in the Karhunen-Loève basis, satisfies the relation,

\[
\int_{\Omega} d\mathbf{s} dt \ \mathbb{E}[|u(\mathbf{s},t;\alpha)- \bar{u}(\mathbf{s},t)|^2] = \sum^{M}_{i=1} \lambda_i \int_{\Omega} d\mathbf{s} dt \ |\phi_i(\mathbf{s},t)|^2 = \sum^{M}_{i=1} \lambda_i = \operatorname{Tr}(C_M) \footnote{Here, $M$ denotes the number of the
truncated spectral expansion. The variance of the associated truncated
field is related to the covariance trace,
$\operatorname{Tr}(C_M)=\sum_{i=1}^{M}\lambda_i$.},
\]

where the cross-terms cancel out due to the independence of the coefficients \(\xi_i(\alpha)\) and the orthogonality of the eigenfunctions \(\phi_i\). \(\square\)

\subsection{Theorem 4: Formal correspondence between the Bayesian likelihood and the Heisenberg observable}
\label{subsec:teo_4}

\textit{Under additive Gaussian observational noise, the quadratic form appearing in the Bayesian likelihood can be expressed in terms of the initial observable of the Heisenberg representation, establishing a formal correspondence between the Bayesian likelihood and the Heisenberg observable structure.}

\subsubsection{Proof}
The Bayesian likelihood is given by \cite{tarantola_2005},

\[
\mathcal{L}(\boldsymbol{\nu}) \propto \exp\left\{ -\frac{1}{2} \| \mathbf{d}^{obs} - \mathbf{f}(\boldsymbol{\nu}) \|_{\Gamma_{\mathrm{noise}}^{-1}}^2 \right\}.
\]

Comparing the residual between the observational noisy data \(\mathbf{d}^{obs}\) and the output of the forward model \(\mathbf{f}(\boldsymbol{\nu}) \), it holds that the residual \( r = \mathbf{d}^{obs} - \mathbf{f}(\boldsymbol{\nu})\) is represented in the KL basis by the residual state \(|\Psi^{obs}\rangle - |\Psi(\boldsymbol{\nu})\rangle\), where \(\boldsymbol{\nu}\) are the parameters of the inverse problem. Furthermore, by the identification established in \ref{sec:bayes}, \(\Gamma^{-1}_{\mathrm{noise}} \equiv \hat{A}(\mathbf{s}, 0)\), the likelihood function in terms of the quantum state is written as,

\[
\mathcal{L}(\boldsymbol{\nu}) \propto \exp\left\{ -\frac{1}{2} \| \Psi^{obs} - \Psi(\boldsymbol{\nu}) \|_{\hat{A}(\mathbf{s}, 0)}^2 \right\},
\]

where the residual \((\Psi^{obs} - \Psi(\boldsymbol{\nu}))\) is weighted by the initial observable \(\hat{A}(\mathbf{s},0)\).

Expanding the quadratic term in the previous expression,
\[
\begin{aligned}
\langle \Psi^{obs} - \Psi(\boldsymbol{\nu}) | \hat{A}(\mathbf{s},0) | \Psi^{obs} - \Psi(\boldsymbol{\nu}) \rangle
&= \langle \Psi^{obs} | \hat{A}(\mathbf{s},0) | \Psi^{obs} \rangle \\
&\quad - 2 \Re \langle \Psi^{obs} | \hat{A}(\mathbf{s},0) | \Psi(\boldsymbol{\nu}) \rangle \\
&\quad + \langle \Psi(\boldsymbol{\nu}) | \hat{A}(\mathbf{s},0) | \Psi(\boldsymbol{\nu}) \rangle.
\end{aligned}
\]

The first term \(\langle \Psi^{obs} | \hat{A}(\mathbf{s},0) | \Psi^{obs} \rangle\) is independent of \(\boldsymbol{\nu}\) and can be absorbed into the proportionality constant. Therefore, the likelihood becomes,
\[
\mathcal{L}(\boldsymbol{\nu}) \propto \exp\left\{ -\frac{1}{2} \left( - 2 \Re \langle \Psi^{obs} | \hat{A}(\mathbf{s},0) | \Psi(\boldsymbol{\nu}) \rangle + \langle \Psi(\boldsymbol{\nu}) | \hat{A}(\mathbf{s},0) | \Psi(\boldsymbol{\nu}) \rangle \right) \right\}.
\]

Assuming a Gaussian prior probability \(\mathcal{L}_{prior}(\boldsymbol{\nu}) \propto \exp\left\{ -\frac{1}{2} \| \boldsymbol{\nu} - \boldsymbol{\nu}_0 \|_{\Gamma_{prior}^{-1}}^2 \right\}\), where \(\Gamma_{prior}\) is the initial covariance of the unknown parameters \(\boldsymbol{\nu}\), and multiplying  the likelihood by the prior probability, the posterior probability is obtained as,

\[
f_{post}(\boldsymbol{\nu}) \propto \exp\left\{ -\frac{1}{2} \left( - 2 \Re \langle \Psi^{obs} | \hat{A}(\mathbf{s},0) | \Psi(\boldsymbol{\nu}) \rangle + \langle \Psi(\boldsymbol{\nu}) | \hat{A}(\mathbf{s},0) | \Psi(\boldsymbol{\nu}) \rangle \right) - \frac{1}{2} \| \boldsymbol{\nu} - \boldsymbol{\nu}_0 \|_{\Gamma_{prior}^{-1}}^2 \right\},
\]

thus, the Bayesian update of the posterior probability is formally expressed in terms of the same weighted quadratic form associated with
the initial observable \(\hat{A}(\mathbf{s},0)\), which in the Heisenberg picture evolves as \(\hat{A}(\mathbf{s},\tau) = U^\dagger(\tau) \hat{A}(\mathbf{s},0) U(\tau)\). 

The cross term in the above expression: \(-2\Re\langle\Psi^{obs}|\hat{A}(\mathbf{s},0) \Psi(\boldsymbol{\nu})\rangle\) depends on the weighted inner product between the observational state and the forward model state, while the second term: \(\langle\Psi(\boldsymbol{\nu})|\hat{A}(\mathbf{s},0)|\Psi(\boldsymbol{\nu})\rangle\) retains the weighted quadratic structure associated with the initial observable.

Thus, the Bayesian posterior \(f_{post}(\boldsymbol{\nu})\) contains the same weighted quadratic structure associated with the initial observable, together with a cross term that
couples the observed and the forward model states. 

Therefore, a formal correspondence is established between the quadratic structure appearing in the Bayesian likelihood and the expectation value associated with the observable in the Heisenberg representation. \(\square\)

\subsection{Limit case \texorpdfstring{\(\ell \to 0\)}{l to 0}}
\label{subsec:limite1}

Next, the limit cases for the Matérn covariance are analyzed. The first case is when the parameter \(\ell \to 0\) in \(C(\mathbf{s}, \mathbf{s'}, t, t')\).

The Matérn covariance of the stochastic equation with white noise given by expression \eqref{eq:edp_white_noise} is written as,

\begin{equation}
C(\mathbf{s}, \mathbf{s'}, t, t') = \frac{2^{1-\eta}}{\Gamma(\eta)} \left(\frac{||\mathbf{s} - \mathbf{s'}||}{\ell}\right)^{\eta} K_{\eta} \left(\frac{||\mathbf{s} - \mathbf{s'}||}{\ell}\right) \frac{2^{1-\eta_{t}}}{\Gamma(\eta_{t})} \left(\frac{|t-{t'}|}{\ell}\right)^{\eta_{t}}  K_{\eta_{t}}\left(\frac{|t-{t'}|}{\ell}\right),
\label{eq:covariance_matern2}
\end{equation}

however, the limit case is considered where the length parameter tends to zero, the argument of the second-order modified Bessel function \(\frac{||\mathbf{s}-\mathbf{s'}||}{\ell}\) tends to infinity, that is, \(\frac{||\mathbf{s}-\mathbf{s'}||}{\ell} \to \infty\), then the second-order modified Bessel function with this argument \cite{abramowitz_1972},

\begin {equation}
K_{\eta}\left(\frac{||\mathbf{s}-\mathbf{s'}||}{\ell}\right) = \frac{\pi}{2} \frac{I_{-\eta}\left(\frac{||\mathbf{s}-\mathbf{s'}||}{\ell}\right) - I_{\eta}\left(\frac{||\mathbf{s}-\mathbf{s'}||}{\ell}\right)}{\sin(\pi\eta)},
\label{eq:bessel_second_order}
\end{equation}
where \( I_{-\nu}\) and \(I_{\nu}\) are first-order modified Bessel functions.

Using the asymptotic behavior as \(\frac{||\mathbf{s}-\mathbf{s'}||}{\ell} \to \infty\) in expression \eqref{eq:bessel_second_order},

\begin{equation}
K_{\eta}\left(\frac{||\mathbf{s}-\mathbf{s'}||}{\ell}\right) \sim \sqrt{\frac{\pi}{2(\frac{||\mathbf{s}-\mathbf{s'}||}{\ell})}} \exp{-\left(\frac{||\mathbf{s}-\mathbf{s'}||}{\ell}\right)} \quad,\ell \to 0.
\label{eq:bessel_second_order2_spatial}
\end{equation}

The function \( K_{\eta}\left(\frac{||\mathbf{s}-\mathbf{s'}||}{\ell}\right)\) in expression \eqref{eq:bessel_second_order2_spatial} decays exponentially to zero, therefore the spatial part of the Matérn covariance tends to zero for \(\|\mathbf{s}-\mathbf{s'}\| \neq 0\).

Similarly, for the time component, as \( \ell \to 0 \), the argument \( \frac{|t-t'|}{\ell} \) tends to infinity for \( |t-t'| \neq 0 \). Using the same asymptotic behavior of the modified Bessel function,

\begin{equation}
K_{\eta_t}\left( \frac{|t-t'|}{\ell} \right) \sim
\sqrt{\frac{\pi}{2 \frac{|t-t'|}{\ell}}}
\exp\left( -\frac{|t-t'|}{\ell} \right), \quad \ell \to 0,
\label{eq:bessel_second_order2_temporal}
\end{equation}

it is found that the temporal component of the covariance tends to zero for \( |t-t'| \neq 0 \),

Combining both spatial and temporal limits yields,

\begin{equation}
C(\mathbf{s}, \mathbf{s'}, t, t')\to 0 \qquad
(\mathbf{s},t)\neq(\mathbf{s'},t').
\label{eq:cov_l0}
\end{equation}

\subsection{Limit cases \texorpdfstring{\(\ell \to \infty\)}{l to infinity}}
\label{subsec:limite2}

In the limit case for the Matérn covariance when \( \ell \to \infty\), the arguments of the second-order modified Bessel functions \( \frac{||\mathbf s - \mathbf{s'}||}{\ell}\) and \( \frac{|t-t'|}{\ell}\) tend to zero. The asymptotic behavior of the spatial Bessel function is,

\begin{equation}
K_{\eta} \sim \frac{\Gamma(\eta)}{2} \left(\frac{2}{\frac{||\mathbf s - \mathbf{s'}||}{\ell}}\right)^{\eta}, \quad \ell \to \infty
\label{eq:bessel_l_infy_espacial}
\end{equation}

The asymptotic behavior of the temporal Bessel function,

\begin{equation}
K_{\eta_{t}} \sim \frac{\Gamma(\eta_{t})}{2} \left(\frac{2}{\frac{|t - t'|}{\ell}}\right)^{\eta_{t}}, \quad \ell \to \infty
\label{eq:bessel_l_infy_temporal}
\end {equation}

Therefore, substituting the asymptotic behavior of the modified Bessel functions of the second kind as \( \ell \to \infty\) into the Matérn covariance \( C(\mathbf{s,s'},t,t')\) yields the following expression,

\begin{equation}
C(\mathbf{s}, \mathbf{s'}, t, t') = \frac{2^{1-\eta}}{\Gamma(\eta)} \left(\frac{||\mathbf{s} - \mathbf{s'}||}{\ell}\right)^{\eta}
\frac{\Gamma(\eta)}{2} \left(\frac{2}{\frac{||\mathbf s - \mathbf{s'}||}{\ell}}\right)^{\eta}\frac{2^{1-\eta_{ t}}}{\Gamma(\eta_{ t})} \left(\frac{|t-{t'} |}{\ell}\right)^{\eta_{t}} \frac{\Gamma(\eta_{t})}{2} \left(\frac{2}{\frac{|t - t'|}{\ell}}\right)^{\eta_t},
\label{eq:covariance_matern3}
\end{equation}

simplifying the expression \eqref{eq:covariance_matern3} for the covariance \( C(\mathbf{s,s'},t,t')\),

\begin{equation}
C(\mathbf{s}, \mathbf{s'}, t, t') = 1,
\end{equation}
that is, the covariance tends to a constant equal to $1$, so when \(\ell \to \infty\), the covariance is not invertible, and the only possible nonzero eigenvalue is \( \lambda_{1}\), which corresponds to a constant mode. 

\section{Numerical examples}
\label{sec:numerical_examples}
In this section, three numerical examples are presented that illustrate and support the analytical results of the preceding theorems.

\subsection{Example 1: Trace invariance and comparison of eigenvalues of \texorpdfstring{\(\hat A(\mathbf s,0)\) and \(\hat A(\mathbf s,\tau)\)}{A(s,0) and A(s,tau)}}
\label{subsec:numerical1}

The goal of this example is to illustrate numerically that the trace of the observable operator, \(\operatorname {Tr}(\hat A(\mathbf s, \tau))\), is invariant under time evolution, and that the eigenvalues of the operators \(\hat A(\mathbf s, 0)\) and \(\hat A(\mathbf s, \tau)\) are conserved, that is, \(\lambda_{\hat A(\mathbf s, 0)} = \lambda_{\hat A(\mathbf s, \tau)}\).

To demonstrate this numerically, first, an initial observable operator \(\hat A(\mathbf s, 0)\) is constructed, which is symmetric and positive definite. This is achieved through the spectral decomposition \(\hat A(\mathbf s, 0) = Q \Lambda Q^{T}\), where \(Q\) is an orthogonal matrix and \(\Lambda\) is a diagonal matrix with positive eigenvalues.

\begin{figure}[ht]
    \centering
    \includegraphics[width=0.5\textwidth]{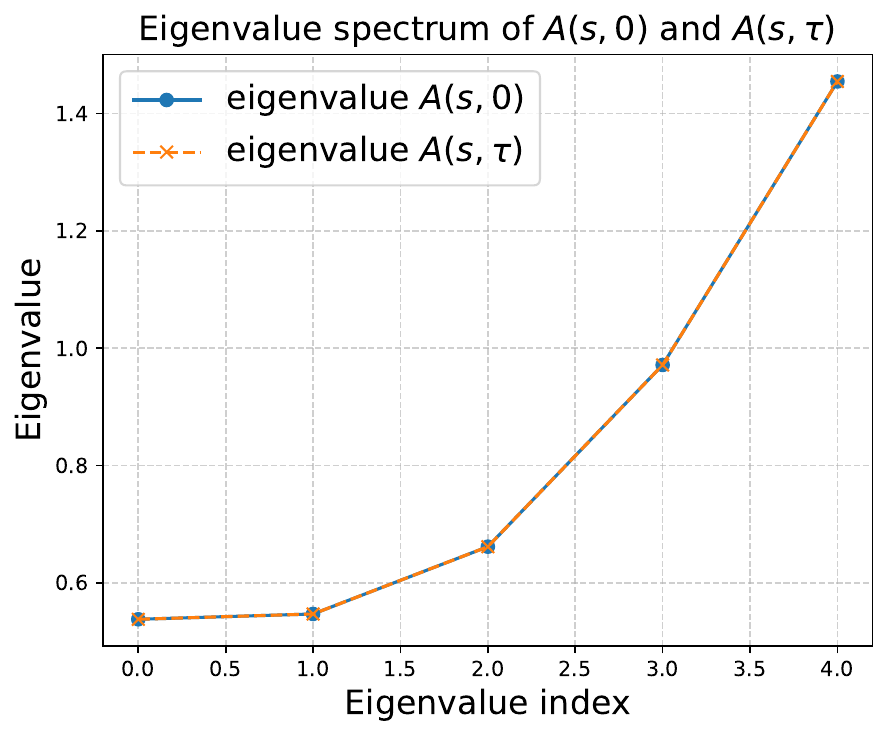}
    \caption{Eigenvalue spectrum of \(\hat A(\mathbf s, 0)\) and \(\hat A(\mathbf s, \tau)\). The overlap shows that the eigenvalues are invariant under unitary time evolution.}
    \label{fig:exp1_spectrum}
\end{figure}

Next, the evolved observable operator \(\hat A(\mathbf s, \tau) = U^{\dagger} \hat A(\mathbf s, 0) U\) is constructed. For the numerical construction of \(U\), a Hermitian Hamiltonian is used, which is independent of the evolution parameter \(\tau\): \(U(\tau)=\exp(-iH\tau)\). Once both operators have been constructed, a spectral decomposition is performed to obtain the eigenvalues \(\lambda_{\hat A(\mathbf s, 0)}\) and \(\lambda_{\hat A(\mathbf s, \tau)}\), and with these results, the graph in figure (\ref{fig:exp1_spectrum}) is plotted, where the spectra of the eigenvalues of the initial observable operator (blue line) and the evolved operator (orange line) are compared. It can be seen that both spectra overlap, with no differences observed, numerically illustrating the conservation of the eigenvalues.

\begin{figure}[ht]
    \centering
    \begin{subfigure}[b]{0.48\textwidth}
        \centering
        \includegraphics[width=\textwidth]{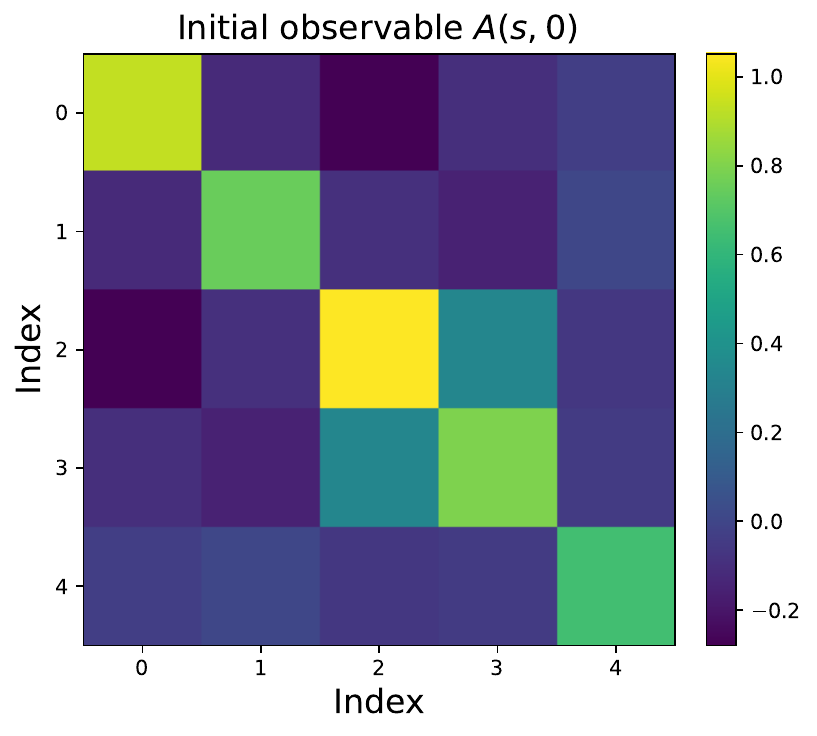}
        \caption{Heat map of the \(\hat A(\mathbf s, 0)\).}
        \label{fig:exp1_A0}
    \end{subfigure}
    \hfill  
    \begin{subfigure}[b]{0.48\textwidth}
        \centering
        \includegraphics[width=\textwidth]{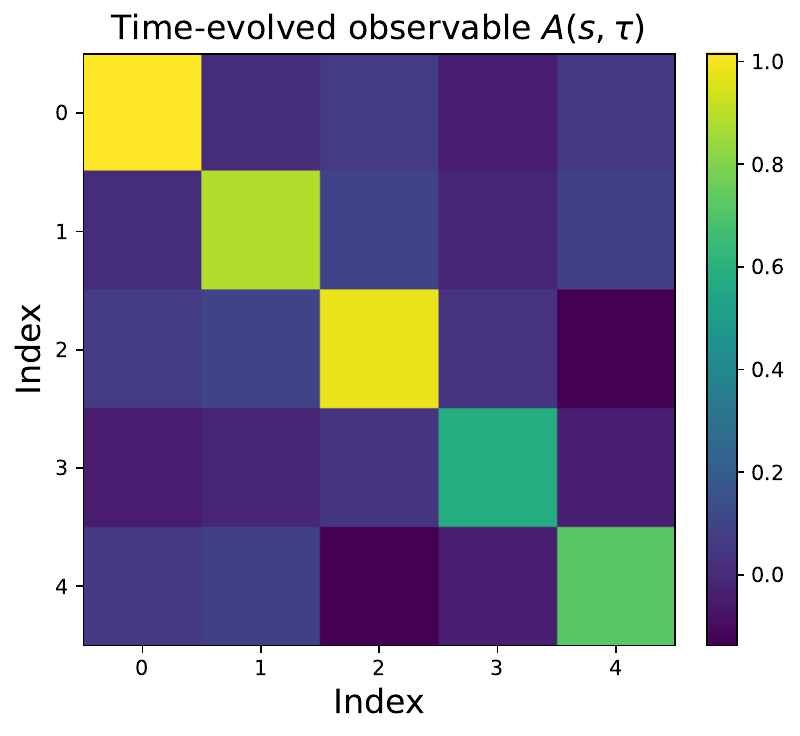}
        \caption{Heat map of the \(A(\mathbf s, \tau)\).}
        \label{fig:exp1_At}
    \end{subfigure}  
    \caption{Comparison of the initial and time-evolved observable matrices.}
    \label{fig:exp1_compare}
\end{figure}

The heat maps of each operator are compared: figure \ref{fig:exp1_A0} shows the heat map for the initial operator \(\hat A(\mathbf s,0)\) and figure \ref{fig:exp1_At} shows the heatmap for the evolved operator \(\hat A(\mathbf s, \tau)\). These heat maps show that the matrix structures of \(\hat A(\mathbf s, \tau)\) and \(\hat A(\mathbf s, 0)\) remain very similar. Furthermore, it has been verified that both traces are conserved: \(\operatorname{Tr}(\hat A(\mathbf s, 0)) = 4.173259\) and \(\operatorname{Tr}(\hat A(\mathbf s, \tau)) = 4.173259\). This conservation follows from the unitarity of the evolution operator (\(U^\dagger U = \hat{I}\)) and provides a numerical illustration of Theorem 1 (unitarity) and Theorem 3 (trace conservation).

\subsection{\texorpdfstring{
Example 2:\(\int_{\Omega}\mathbb{E}[|u(\mathbf{s},t;\alpha)|^2]
\ d\mathbf{s}\ dt=\sum_i \lambda_i=\operatorname{Tr}[C]\)}{Example 2: Total variance and trace of the covariance operator}}
\label{subsec:numerico2}

The goal of this example is to show that the expected value of the square of a stochastic field expanded using the KL expansion can be expressed as the trace of the operator. Since a zero mean stochastic field is considered in this example, that is, \(\bar{u}(\mathbf s,t)=0\), the following relation holds:  \(\int_{\Omega}\mathbb{E}[|u(\mathbf{s},t;\alpha) |^2]\ d\mathbf{s}\ dt = \sum_{i} \lambda_{i} = \operatorname{Tr}[{C}]\), thereby, illustrating that the total variance of the stochastic field is given by the trace of the covariance operator. The field is expressed as,
\[
u(\mathbf{s},t;\alpha) = \sum_{i=1}^{\infty} \sqrt{\lambda_i} \ \phi_i(\mathbf{s},t) \ \xi_i(\alpha),
\]
where \(\xi_{i}(\alpha)\) are independent random variables with mean zero and unit variance.

To perform the numerical demonstration, the Matérn covariance is first constructed, which is given by expression \eqref{eq:covariance_matern}, with parameters \(\eta=\eta_{t}=1.5\) and \(\ell_{s}=\ell_{t}=0.2\) for the spatial and temporal components. For this numerical example, a one-dimensional spatial domain \(D=[0,1]\subset\mathbb{R}\) and a temporal domain \(I=[0,1]\) are considered. The spatial and temporal domains are each discretized using \(10\) points, resulting in a total of \(100\) space-time points. This grid is a very coarse discretization, but is sufficient for the purpose of numerically illustrating the  results.

Once the Matérn covariance is obtained, a spectral decomposition is performed to obtain the eigenvalues and eigenfunctions. Using the KL expansion, the field is expanded according to expression \eqref{eq:KL_expantion}, and the total variance is then obtained using the expected value expression \(\int_{\Omega}\mathbb{E}[|u(\mathbf{s},t; \alpha)|^2] d\mathbf{s}\ dt\) and the sum of the eigenvalues \(\sum_{i} \lambda_{i} \). The KL discrete expansion with $100$ modes is used, and $500$ independent realizations are generated. The total variance of the field is found to be \(\int_{\Omega}\mathbb{E}[|u(\mathbf{s},t;\alpha)|^2]\ d\mathbf{s}\ dt = 96.5491\), while the sum of the eigenvalues is \(\sum_{i} \lambda_{i} = 100.0000 \). These two values differ by approximately \(3.45\) (a percentage difference of $3.45\%$), which can be attributed to the finite number of stochastic realizations used to approximate the expectation.

\begin{figure}[ht]
    \centering
\includegraphics[width=0.6\textwidth]{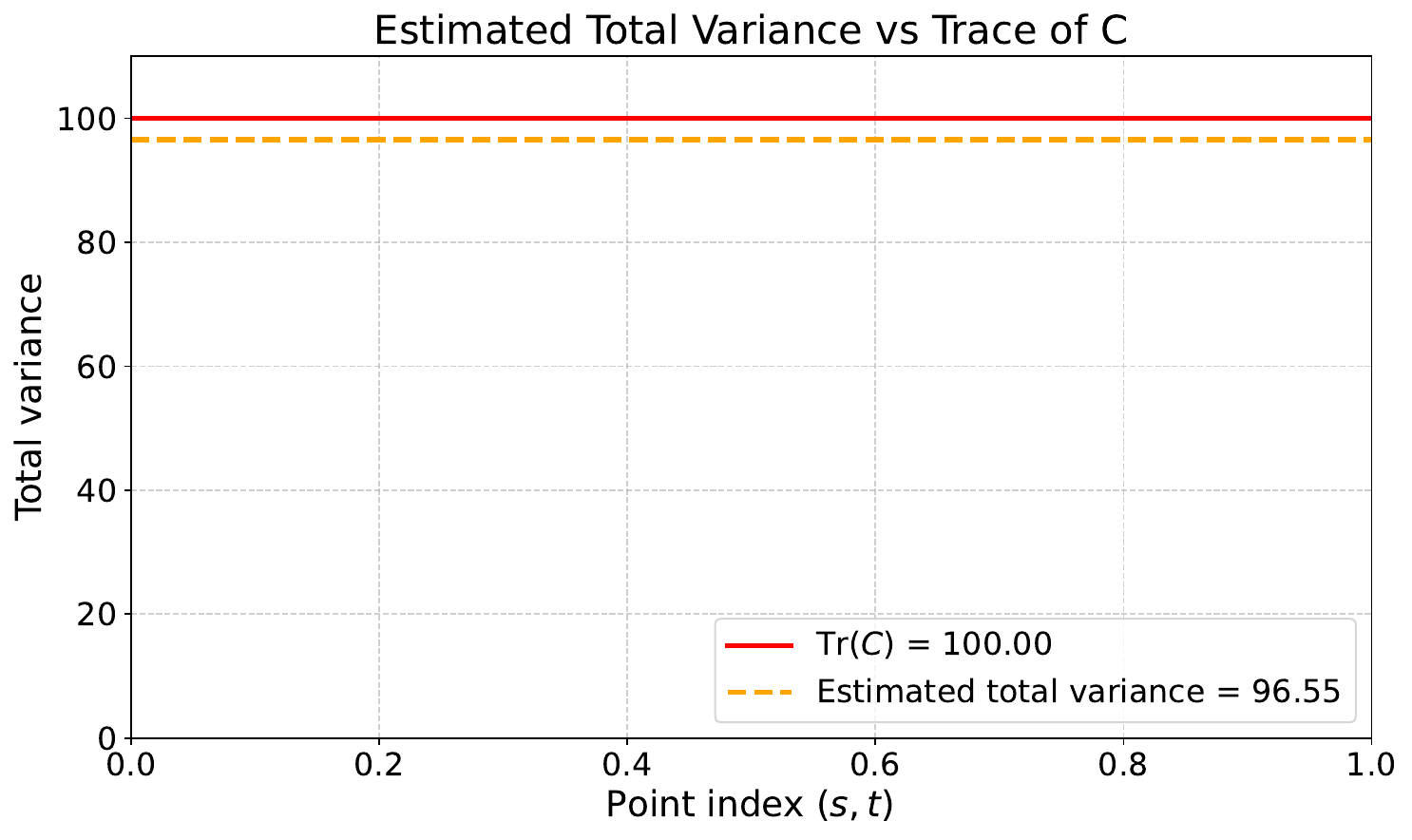}
    \caption{Comparison between the total variance of the system \(\int_{\Omega}\mathbb{E}[|u(\mathbf{s},t;\alpha)|^2]\ d\mathbf{s}\ dt\) and the trace of the operator \(\operatorname{Tr}[{C}]\).}
    \label{fig:exp2_variance}
\end{figure}

Figure \ref{fig:exp2_variance} shows a comparison between the total variance (orange line) and the trace of the covariance \(\operatorname{Tr}[C]\) (red line). The closeness of the two lines is consistent with the identity \(\int_{\Omega}\mathbb{E}[|u(\mathbf{s},t;\alpha)|^2]\ d\mathbf{s}\ dt = \sum_{i}\lambda_{i}=\operatorname{Tr}[{C}]\).

\newpage

\subsection{Example 3: Numerical verification of the Bayesian quadratic form in the KL basis}
\label{subsec:numerico3}

This numerical example verifies the equivalence between the Bayesian quadratic form evaluated in the original representation and its representation in the KL basis, derived in appendix \ref{app:appendixA}. A synthetic forward-model response \(f(\boldsymbol{\nu})\) is generated using a sinusoidal function depending on the spatial and temporal coordinates. Gaussian observational noise is then added to obtain the observational data \(\mathbf d^{obs}\). 

The KL eigenfunctions \(\phi_i\) are obtained from the spectral decomposition of the Matérn covariance operator introduced in section \ref{subsec:kl}. The parameters considered are \(\ell_s=\ell_t=1\) and \(\eta=\eta_t=0.5\), for which the covariance reduces to the product of exponential correlations given in \eqref{eq:cov_matern_exp}. The same space-time discretization as in example 2 is considered, with a  one-dimensional spatial domain \(D=[0,1]\subset\mathbb{R}\) and a temporal domain \(I=[0,1]\).

Following the formulation introduced in section \ref{sec:bayes}, the residual is \(r(\boldsymbol{\nu}) =\mathbf d^{obs}-\mathbf f(\boldsymbol{\nu}),\) and the squared weighted norm appearing in the exponent of the likelihood,

\begin{equation}
J = r^{T}\Gamma_{\mathrm{noise}}^{-1} r.
\label{eq:J_data}
\end{equation}

To evaluate the same quadratic form in the KL representation, the residual is projected onto the KL eigenfunctions, that is,

\begin{equation}
r_i(\boldsymbol{\nu}) = \langle\phi_i| r(\boldsymbol{\nu})\rangle.
\label{eq:r_KL_num}
\end{equation}

For the numerical verification, the square weighted norm \(J\) is evaluated in the original representation, while \(J_{\mathrm{KL}}^{(M)}\) denotes the corresponding quantity evaluated after projecting the residual into the first \(M\) KL modes. 

The observational noise considered in this example is assumed to be uncorrelated and Gaussian, with \(\Gamma_{\mathrm{noise}}=\sigma_{\mathrm{noise}}^2 I\). According to the derivation presented in appendix \ref{app:appendixA}, when the complete KL basis is considered, that is \(M=N\), both representations are equivalent \(J=J_{\mathrm{KL}}^{(N)}\). To evaluate the effect of the KL truncation, the relative error is defined as,

\begin{equation}
e_{\text{rel}}^{(M)} = \frac{|J-J_{\mathrm{KL}}^{(M)}|
}{|J|}.
\label{eq:relative_error_KL}
\end{equation}

\begin{figure}[ht]
    \centering
    \includegraphics[width=0.6\textwidth]{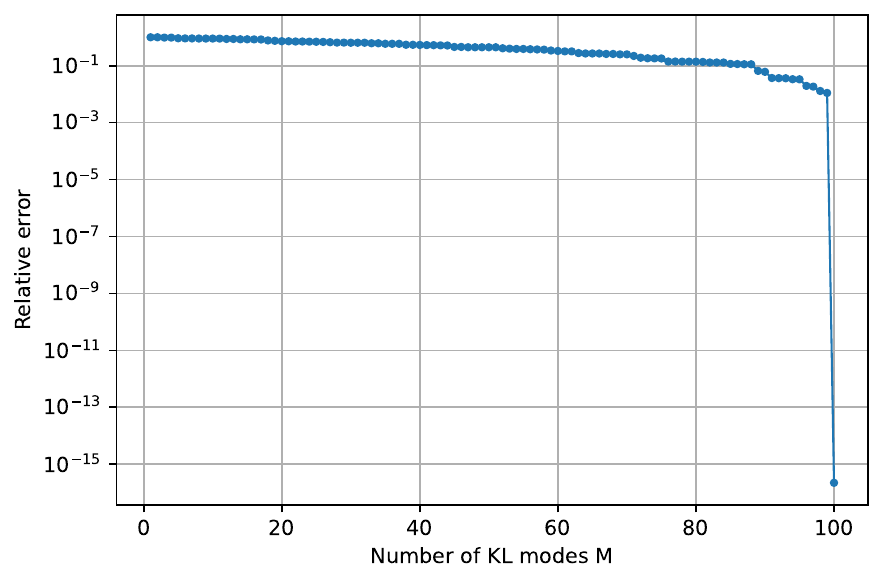}
    \caption{Relative error between the Bayesian quadratic form evaluated in the original data representation and its truncated KL representation as a function of the number of KL modes.}
    \label{fig:exp3_KL_convergence}
\end{figure}

Figure \ref{fig:exp3_KL_convergence} shows the convergence of the relative error as the number of KL modes increases up to the complete discretized basis, that is \(M=N=100\) terms.

It can be seen that the relative error between the two quadratic forms decreases as the number \(M\) in the KL expansion increases. When the complete basis is considered, \(M= N= 100\), the relative error decreases to \(e_{\text{rel}}= 1.37 \times 10^{-16}\), confirming the equivalence derived in appendix \ref{app:appendixA}.

\section{Conclusions}
\label{sec:conclusions}

In this work, a formal correspondence between a Bayesian inverse problem and the Heisenberg representation in quantum mechanics has been formulated. The central focus of this formulation has been the identification of the initial observable operator with the inverse observational noise covariance operator appearing in the likelihood function. To establish this formulation, a stochastic field with Matérn covariance was represented through the Karhunen–Loève expansion, and the eigenvalues and eigenfunctions of this expansion were used to construct the state \(|\Psi\rangle\). This formalism is completed by the construction of a Fock space and a self-adjoint Hamiltonian (as demonstrated in theorem 2), whose unitary evolution governs the evolution of the observable, as shown in theorem 1.

Four theorems were presented that establish the mathematical properties of the proposed formalism: the time-evolution operator is unitary, the Hamiltonian is self-adjoint, the trace of the observable is invariant under unitary evolution and therefore conserved, and the quadratic form appearing in the Bayesian likelihood is formally related to the observational operator. Based on this mathematical correspondence, the initial observable in the Heisenberg representation is proposed as the inverse observational noise covariance operator. Furthermore, an analysis of the limiting cases of the Matérn covariance was performed as \(\ell\to0\) and \(\ell\to\infty\),  corresponding to vanishing correlations between distinct space-time points and a constant mode, respectively. 

The analytical results were illustrated through three numerical examples. The first example verifies the invariance of the trace of the observable during the evolution in \(\tau\) and the preservation of its eigenvalues under unitary evolution, illustrating the results of theorems 1 and 3. The second example illustrates the relation between the trace of the Matérn covariance operator and the total variance represented by the KL expansion, obtaining a relative difference of \(3.45\%\). Finally, the third example compares the quadratic form evaluated in the original data representation with the corresponding representation in the KL basis. When the complete KL is considered, that is \(M=N=100\), a relative error \(1.37\times10^{-16}\) is obtained between the two representation, numerically illustrating the correspondence established in theorem 4.

This work offers a new perspective on the treatment of inverse problems under uncertainty, combining concepts from stochastic field theory and the Heisenberg representation. The correspondence developed here is formal and does not imply the quantization of the underlying physical inverse problem. The proposed correspondence could provide a basis for future extensions, such as the study of nonlinear forward models. An additional direction for future work is to investigate whether the proposed operator framework can be used to develop alternative strategies for characterizing unknown observational noise covariance operators.

\section{Acknowledgments}

The author acknowledges the Facultad de Ingeniería at Universidad Alberto Hurtado for its research funding.

\appendix
\setcounter{section}{0}
\section{Representation of the Bayesian quadratic form in the KL basis}
\label{app:appendixA}

The correspondence between the quadratic likelihood and Heisenberg representation, requires the likelihood projection into KL eigenfunction. As defined before, the residual between the forward model response and the observational data can be written as,

\begin{equation}
r(\boldsymbol{\nu})=
\mathbf{d}^{obs}-\mathbf{f}(\boldsymbol{\nu}).
\end{equation}

The observational data and the forward-model response can also be represented in the KL basis as, 

\begin{equation}
|\Psi^{obs}\rangle=\sum_i d_i^{obs}|\phi_i\rangle,\qquad |\Psi(\boldsymbol{\nu})\rangle = \sum_i f_i(\boldsymbol{\nu})|\phi_i\rangle,
\end{equation}

where \(d_i^{obs}\) and \(f_i(\boldsymbol{\nu})\) are the KL coefficients obtained by projecting \(\mathbf{d}^{obs}\) and \(\mathbf{f}(\boldsymbol{\nu})\), respectively, into the eigenfunctions \(\phi_i\). Therefore, the residual state in KL basis is,

\begin{equation}
|\Psi_r(\boldsymbol{\nu})\rangle = \sum_i r_i |\phi_i\rangle, \qquad r_i=d_i^{obs}-f_i(\boldsymbol{\nu}).
\end{equation}

Using the identification \(\hat{A}(\mathbf{s},0)=\Gamma_{\mathrm{noise}}^{-1}\), the quadratic form can be written as

\begin{align}
\langle\Psi_r|\hat{A}(\mathbf{s},0)|\Psi_r\rangle &= \left(\sum_i r_i^{*}\langle\phi_i|\right) \Gamma_{\mathrm{noise}}^{-1} \left(\sum_j r_j|\phi_j\rangle\right)\\&= \sum_{i,j}r_i^{*}r_j \langle\phi_i| \Gamma_{\mathrm{noise}}^{-1}|\phi_j\rangle.
\end{align}

For the case of uncorrelated Gaussian observational noise, the covariance operator and its inverse are given by,

\begin{equation}
\Gamma_{\mathrm{noise}} = \sigma_{\mathrm{noise}}^2 I, \qquad \Gamma_{\mathrm{noise}}^{-1} = \frac{1}{\sigma_{\mathrm{noise}}^2}I.
\end{equation}

Therefore,

\begin{align}
\langle\Psi_r|\hat{A}(\mathbf{s},0)|\Psi_r\rangle &= \frac{1}{\sigma_{\mathrm{noise}}^2} \sum_{i,j} r_i^{*}r_j \langle\phi_i|\phi_j\rangle\\
&=\frac{1}{\sigma_{\mathrm{noise}}^2}\sum_{i,j} r_i^{*}r_j\delta_{ij}\\&= \frac{1}{\sigma_{\mathrm{noise}}^2}\sum_i |r_i|^2.
\end{align}

Since the KL eigenfunctions are an orthonormal basis,

\begin{equation}
\sum_i |r_i|^2= r^{\dagger}r.
\end{equation}

Consequently,

\begin{align}
\langle\Psi_r|\hat{A}(\mathbf{s},0)|\Psi_r\rangle &= \frac{1}{\sigma_{\mathrm{noise}}^2}{r}^{\dagger}{r}\\ &= (\mathbf{d}^{obs}-\mathbf{f}(\boldsymbol{\nu}))^{\dagger}\frac{1}{\sigma_{\mathrm{noise}}^2}I(\mathbf{d}^{obs}-\mathbf{f}(\boldsymbol{\nu}))\\
&=(\mathbf{d}^{obs}-\mathbf{f}(\boldsymbol{\nu}))^{\dagger} \Gamma_{\mathrm{noise}}^{-1}(\mathbf{d}^{obs}-\mathbf{f}(\boldsymbol{\nu})),
\end{align}

the above expression, considering only  real observational data and real forward model responses, it follows,

\begin{equation}
\langle\Psi_r|\hat{A}(\mathbf{s},0)|\Psi_r\rangle
=(\mathbf{d}^{obs}-\mathbf{f}(\boldsymbol{\nu}))^{T}
\Gamma_{\mathrm{noise}}^{-1}(\mathbf{d}^{obs}-\mathbf{f}(\boldsymbol{\nu})).
\end{equation}


\begin{thebibliography}{99}

\bibitem{abramowitz_1972}
Abramowitz, M., \& Stegun, I. A. (1972). \textit{Handbook of Mathematical Functions}. Dover Publications.

\bibitem{abugattas_entropy_2026}
Abugattas, C., Carpio, A., \& Cebrián, E. (2026). Uncertainty quantification in inverse scattering problems. \textit{Entropy}, \textit{28}(4), 461.

\bibitem{abugattas_quantifying_2025}
Abugattas, C., Carpio, A., Cebrián, E., \& Oleaga, G. (2025). Quantifying uncertainty in inverse scattering problems set in layered environments. \textit{Applied Mathematics and Computation}, \textit{500}, 129453.

\bibitem{arai_fock}
Arai, A. (2018). \textit{Analysis on Fock Spaces and Mathematical Theory of Quantum Fields: An Introduction to Mathematical Analysis of Quantum Fields}. World Scientific.

\bibitem{aster_2018}
Aster, R. C., Borchers, B., \& Thurber, C. H. (2018). \textit{Parameter Estimation and Inverse Problems}. Elsevier.

\bibitem{attal_fock}
Attal, S. (2015). \textit{Fock spaces}. Lecture notes, Institut Camille Jordan, Université Lyon 1.

\bibitem{Bishop2006}
Bishop, C. M. (2006). \textit{Pattern Recognition and Machine Learning}. Springer.

\bibitem{calvetti_2008}
Calvetti, D., \& Somersalo, E. (2008). Hypermodels in the Bayesian imaging framework. \textit{Inverse Problems}, \textit{24}, 034013.

\bibitem{carpio_2020}
Carpio, A., Iakunin, S., \& Stadler, G. (2020). Bayesian approach to inverse scattering with topological priors. \textit{Inverse Problems}, \textit{36}, 105001.

\bibitem{carpio_2023}
Carpio, A., Cebrián, E., \& Gutiérrez, A. (2023). Object based Bayesian full-waveform inversion for shear elastography. \textit{Inverse Problems}, \textit{39}, 075007.

\bibitem{derezinski_qft}
Dereziński, J. (2020). \textit{Mathematical Introduction to Quantum Field Theory}. Department of Mathematical Methods in Physics, University of Warsaw.

\bibitem{fetter_walecka_2003}
Fetter, A. L., \& Walecka, J. D. (2003). \textit{Quantum Theory of Many-Particle Systems}. Dover Publications.

\bibitem{ghanem_spanos_2003}
Ghanem, R. G., \& Spanos, P. D. (2003). \textit{Stochastic Finite Elements: A Spectral Approach}. Dover Publications.

\bibitem{horn_johnson_2013}
Horn, R. A., \& Johnson, C. R. (2013). \textit{Matrix Analysis}. Cambridge University Press.

\bibitem{janson_distance}
Janson, S. (2021). On distance covariance in metric and Hilbert spaces. \textit{ALEA, Lat. Am. J. Probab. Math. Stat.}, \textit{18}, 1353-1393.

\bibitem{Jaynes2003}
Jaynes, E. T. (2003). \textit{Probability Theory: The Logic of Science}. Cambridge University Press.

\bibitem{kaipio_2005}
Kaipio, J., \& Somersalo, E. (2005). \textit{Statistical and Computational Inverse Problems}. Springer.

\bibitem{levenberg_1944}
Levenberg, K. (1944). A method for the solution of certain non-linear problems in least squares. \textit{Quarterly of Applied Mathematics}, \textit{2}(2), 164-168.

\bibitem{lindgren_2011}
Lindgren, F., Rue, H., \& Lindström, J. (2011). An explicit link between Gaussian fields and Gaussian Markov random fields: the stochastic partial differential equation approach. \textit{Journal of the Royal Statistical Society: Series B}, \textit{73}(4), 423-498.

\bibitem{marquardt_1963}
Marquardt, D. W. (1963). An algorithm for least-squares estimation of nonlinear parameters. \textit{Journal of the Society for Industrial and Applied Mathematics}, \textit{11}(2), 431-441.

\bibitem{momeni_fock}
Momeni, D. (2025). Reconstructing configurational Hamiltonians from Fock space. \textit{Nuclear Physics B}, \textit{1019}, 117119.

\bibitem{nahakara_geometry}
Nakahara, M. (2003). \textit{Geometry, Topology and Physics} (2nd ed.). CRC Press.

\bibitem{peskin_schroeder_1995}
Peskin, M. E., \& Schroeder, D. V. (1995). \textit{An Introduction to Quantum Field Theory}. Westview Press.

\bibitem{reed_simon_1980}
Reed, M., \& Simon, B. (1980). \textit{Methods of Modern Mathematical Physics I: Functional Analysis} (2nd ed.). Academic Press.

\bibitem{rue_held_2005}
Rue, H., \& Held, L. (2005). \textit{Gaussian Markov Random Fields: Theory and Applications}. Chapman \& Hall/CRC.

\bibitem{sakurai_2020}
Sakurai, J. J., \& Napolitano, J. (2020). \textit{Modern Quantum Mechanics} (3rd ed.). Cambridge University Press.

\bibitem{salcedo_qft}
Salcedo, L. L. (2018). \textit{Introducción a la teoría de campos cuánticos}. Departamento de Física Atómica, Molecular y Nuclear, Universidad de Granada.

\bibitem{stuart_2010}
Stuart, A. M. (2010). Inverse problems: A Bayesian perspective. \textit{Acta Numerica}, \textit{19}, 451-559.

\bibitem{tarantola_2005}
Tarantola, A. (2005). \textit{Inverse Problem Theory and Methods for Model Parameter Estimation}. SIAM.

\end{thebibliography}
\end{document}